\documentclass{aa}  

\usepackage{graphicx}
\usepackage{xcolor}
\usepackage{txfonts}
\usepackage{cancel,soul,ulem}
\defcitealias{Leinhardt2012}{LS12}
\begin{document}

   \title{Protoplanet collisions: new scaling laws from SPH simulations}


   \author{S. Crespi
          \inst{1,2}\fnmsep\thanks{\email{sc6459@nyu.edu}}
          \and
          M. Ali-Dib\inst{2}
          \and
          I. Dobbs-Dixon\inst{1,2,3}
          }

   \institute{
        Department of Physics, New York University Abu Dhabi, PO Box 129188, Abu Dhabi, UAE
         \and
        Center for Astrophysics and Space Science (CASS), New York University
Abu Dhabi, PO Box 129188, Abu Dhabi, UAE
         \and
        Center for Space Science, NYUAD Research Institute, New York University Abu Dhabi, PO Box 129188, Abu Dhabi, UAE 
             }

   \date{Received xxxx; accepted yyyy, published zzzz}

 
\abstract{
One common approach for solving collisions between protoplanets in simulations of planet formation is to employ analytical scaling laws.
The most widely used one was developed by \cite{Leinhardt2012} from a catalog of $\sim$180 N-body simulations of {rubble--pile} collisions.
In this work, we use a new catalogue of more than 20,000 SPH simulations to test the validity and the prediction capability of \cite{Leinhardt2012} scaling laws.
We find that these  laws overestimate the fragmentation efficiency in the merging regime and they are not able to properly reproduce the collision outcomes in the super--catastrophic regime.
In the merging regime, we also notice a significant dependence between the collision outcome, in terms of the largest remnant mass, and the relative mass of the colliding protoplanets.
Here, we present a new set of scaling laws that are able to better predict the collision outcome in all regimes and it is also able to reproduce the observed dependence on the mass ratio.
We compare our new scaling laws against a machine learning approach and obtain similar prediction efficiency.
}


   \maketitle
%

\section{Introduction} \label{sec:intro}

Pairwise collisions are considered to be the main mechanism that drives the growth of planetesimals ($\sim 100$ km sized rocky bodies) into terrestrial planets, in particular inside the water snow line (e.g.: \citealt{Wetherill1980}, \citealt{Kokubo1996}, \citealt{Chambers2001}, \citealt{Izidoro2017}).
Including collisions in simulations of terrestrial planet formation, however, has proven to be particularly challenging for two main reasons.
First of all, the typical timescales involved in collisions are orders of magnitude shorter than the orbital period of planetesimals (\citealt{Benz2007}, \citealt{Kegerreis2020}).
Secondly, it is computationally challenging to fully resolve and integrate the evolution of all the material ejected during the collision, which ranges from gaseous material, like the atmosphere and vaporised rock and water, to solid fragments the size of big asteroids (\citealt{Leinhardt2012}, \citealt{Kegerreis2020}, \citealt{Crespi2021}).

\begin{table*}[t]
  \centering
  \begin{tabular}{l|c|c|c|c|c|c}
  Reference & SPH algorithm & $N_\mathrm{sim}$ & $M_\mathrm{t}/\mathrm{M}_\oplus$ & $M_\mathrm{p}/M_\mathrm{t}$ & composition & {EoS}\\
  \hline
  \hline
    \cite{Denman2020} & \texttt{GADGET}$^\dagger$ & 122 & $3.27$, $6.26$, $10.5$ & $0.04-0.92$ & Fe, silicate and H (gas) & MANEOS\\
    \cite{Gabriel2020} & \texttt{SPHLATCH}$^\star$ & 1401 & $6\cdot 10^{-3}-0.9$ & $0.1-0.7$ & Fe, quartz and H$_2$0 & not specified\\
    \cite{Timpe2020} & \texttt{Gasoline}$^\bullet$ & 10662 & $0.05-1.8$ & $0.1-1$ & Fe and granite & Tillotson\\
    \cite{Burger2020} & \texttt{miluphcuda}$^\ast$ & 9980 & $6\cdot 10^{-4}-1.8$ & $7\cdot 10^{-5}-1$ & Fe, silicate and H$_2$0 & Tillotson\\
    \cite{Crespi2021} & \texttt{miluphcuda}$^\ast$ & 858 & $1.6\cdot 10^{-4}-1.8$ & 0.1, 0.5, 1 & 80-90\% basalt, 10-20\% H$_2$0 & Tillotson\\
    \cite{Winter2022} & \texttt{miluphcuda}$^\ast$ & 10164 & $2.6\cdot 10^{-4}-1.9$ & $0.05 - 1$& Fe, silicate and H$_2$0 & Tillotson\\
  \hline
    \cite{Benz2007} & generic SPH$^\ddagger$ & 17 & 0.1375 & 1/10, 1/6, 1/5 & 33\% Fe, 66\% dunite & ANEOS\\
    \cite{Marcus2009} & \texttt{GADGET}$^\dagger$ & $\sim60$ & 1, 5, 10 & 1/4, 1/2, 3/4& 33\% Fe, 66\% fosterite & MANEOS\\
    \cite{Marcus2010} & \texttt{GADGET}$^\dagger$ & $\sim100$& 0.5-5 & 1/4, 1/2, 1& 50\% serpentine, 50\% H$_2$0 & MANEOS\\
  \end{tabular}
  \vspace{3mm}
  \caption{The three catalogues used in this study are listed in the upper part of the table, while the catalogues used in \citetalias{Leinhardt2012} for gravity--dominated bodies are listed in the bottom part. The columns show (from left to right) the reference paper in which the dataset is presented, the SPH algorithm used to perform the simulations, the total number of simulations ($N_\mathrm{sim}$), the sampled masses of the target body in Earth mass ($M_t/\mathrm{M}_\oplus$), the mass of the projectile scaled by the target mass ($M_\mathrm{p}/M_\mathrm{t}$), the composition of the colliding bodies, and {the equation of state (EoS) used in the simulations. In particular, for the EoS column, the following labels are used: \textit{Tillotson} for the nonlinear equations of state formulated by \cite{Tillotson1962}, \textit{ANEOS} for the set of analytical equations of state developed by \cite{Thompson1972}, and \textit{MANEOS} for the set of analytical equations of state that also account for the energetic effects of the formation of molecular clusters, as developed by \cite{Melosh2007} based on the work of \cite{Thompson1972}.}  $^\dagger$\cite{Springel2005}. $^\star$\cite{Reufer2011} and \cite{Emsenhuber2018}.
  $^\bullet$\cite{Wadsley2004}, \cite{Chau2018} and \cite{Reinhardt2020}. {$^\ddagger$\cite{Monaghan1992}.} 
  $^\ast$\cite{Schafer2016} and \cite{Schafer2020}.}
  \label{tab:1}
\end{table*}

The first challenge can be overcome by employing symplectic and hybrid-symplectic integrators such as \texttt{SyMBA} \citep{Duncan1998} and \texttt{MERCURY} \citep{Chambers1999}, which are able to integrate close-encounters and collisions without significantly affecting the integration precision.
The second challenge, however, is still an open subject of study.
Various approaches have been tested in the past couple of decades.
The most simple approach is to consider all the collisions to result in the inelastic merging of the two colliders.
This approximation has been able to efficiently reproduce the fundamental characteristics of the Solar System, proving the importance of the role played by collisions between protoplanets in the evolution of planetary systems (\citealt{Wetherill1994}, \citealt{Chambers1998}, \citealt{Quintana2002}, \citealt{Raymond2004}, \citealt{OBrien2006}, \citealt{Raymond2006}, \citealt{Quintana2016}).
However, recent studies have shown that, when collisions are assumed to be inelastic, the formation timescale of terrestrial planets is significantly reduced, and both the mass and the water content of the final population of planets are overestimated ({\citealt{Chambers2013}, \citealt{Leinhardt2015}, \citealt{Burger2018}, \citealt{Burger2020}}).
On the other hand, the impact of fragmentation on the evolution of terrestrial planets remains a topic of debate, particularly in investigating the formation of close-in planets in the observed population (e.g. \citealt{Mustill2018}, \citealt{Poon2020}, \citealt{Esteves2022}).

A more sophisticated approach is to allow fragmentation during collisions and to estimate the properties of the main post--collisional bodies by resorting to scaling laws {(e.g. \citealt{Chambers2013}, \citealt{Quintana2016}, 
\citealt{Wallace2017}, 
\citealt{Mustill2018}, \citealt{Clement2019a}, \citealt{Clement2019b}, \citealt{Poon2020},
\citealt{Ishigaki2021},
\citealt{Clement2022})
}.
{Utilizing direct N-body simulations of collisions between rubble--pile differentiated protoplanets \citep{Leinhardt2000}}, \cite{Kokubo2010} and \cite{Leinhardt2012}, derived empirical formulae through which to estimate the mass of the main collisional remnants.
In particular, the widely used scaling laws from the pioneering work of \cite{Leinhardt2012}, hereafter LS12, allow one to directly estimate the mass of the first and second largest remnant given the collision properties.
However, the dataset used to derive the \citetalias{Leinhardt2012} scaling laws, which counts around 180 simulations, was limited to a total of 23 datapoints in the mass range \mbox{$10^{-3}-10$ M$_\oplus$}, of which only 3 datapoints were in the super--catastrophic regime.

Over the past few years, extensive catalogues of {Smooth Particle Hydrodynamics\footnote{For more details about Smooth Particle Hydrodynamics (SPH) simulations we refer the reader to the works of \cite{Benz1990} and \mbox{\cite{Monaghan1992}}.} (SPH) simulations of collisions} have been performed and are now available.
{In particular, \cite{Burger2020} conducted a series of 48 simulations to explore terrestrial planet formation, incorporating on-the-fly SPH simulations to model collisions. This effort yielded a collection of 9,980 simulations of collision between protoplanets.} 
Additionally, \cite{Winter2022} increased by more than ten-fold their previous catalogue of 858 SPH simulations (presented in \citealt{Crespi2021}) by conducting 10,164 new simulations, which also accounted for the rotational momentum of the colliding bodies.

In this work, we make use of these new catalogues of SPH simulations to test and to improve the \citetalias{Leinhardt2012} scaling laws. We propose a new version of their model that is more {accurate} in predicting the mass of the largest post--collisional remnant.
Furthermore, we validate this new set of scaling laws by comparing its prediction efficiency against a machine learning approach.
In Section 2, we present (i) the new best fit parameters for \citetalias{Leinhardt2012} scaling laws, (ii) a new set of scaling laws, and (iii) the machine learning model used to test and validate the new scaling law.
The last section is devoted to the discussion of our results and the main conclusions.

\section{Improved fits to the new SPH dataset} \label{sec:methods}

The original dataset used by \citetalias{Leinhardt2012} to derive their scaling law spans from kilometer-sized bodies to 10 Earth-mass bodies.
Thanks to this wide range of masses, the authors were able to observe a transition between collisions involving small weak bodies and collisions between larger gravity--dominated bodies, with a transition point around $\sim10^{-2}$ M$_\oplus$.
While the scaling laws are the same in the two regimes, the parameters that govern the scaling law differ.
Here, we decided to focus on collision between gravity--dominated bodies only, since the new dataset of SPH simulations used in this study mainly involve gravity--dominated bodies.
We refer the reader to the works of \cite{Burger2020}, \cite{Crespi2021} and \cite{Winter2022} for more details on the three catalogues of SPH simulations.
A summary of these catalogues, together with the datasets used by \citetalias{Leinhardt2012}, is presented in Table \ref{tab:1} for convenience. 

\subsection{Analytical fits}

Based on the model adopted in \citetalias{Leinhardt2012}, the mass of the largest post--collisional remnant ($M_\mathrm{lr}$), scaled by the total mass involved in the collision ($M_\mathrm{tot}$), can be expressed as the function of the relative impact energy ($Q_\mathrm{R}$) scaled by the catastrophic disruption \mbox{criterion ($Q^\star_\mathrm{RD}$)}.
This relation can be written as:
\begin{equation}
\label{eq:universallaw}
    \frac{M_\mathrm{lr}}{M_\mathrm{tot}} =\left\{
    \begin{array}{ll}
        1-0.5\frac{Q_\mathrm{R}}{Q_\mathrm{RD}^\star}
            & \text{  for }  \frac{Q_\mathrm{R}}{Q_\mathrm{RD}^\star}<1.8 \\
        \frac{0.1}{1.8^\eta}\left(\frac{Q_\mathrm{R}}{Q_\mathrm{RD}^\star}\right)^\eta
            & \text{  for }  \frac{Q_\mathrm{R}}{Q_\mathrm{RD}^\star}>1.8
    \end{array}
    \right.
\end{equation}

In the first branch, often referred to as universal law, \citetalias{Leinhardt2012} assume linearity between the impact energy and the largest remnant mass, with the catastrophic disruption criterion ($Q^\star_\mathrm{RD}$) defined as the energy at which half of the total mass is dispersed.
The assumption of linearity, presented also in previous works (\citealt{Stewart2009} and \citealt{Leinhardt2009}), is a good model for collisions with energy close to the catastrophic disruption criterion. However, this assumption actually fails to properly represent collisions with low impact energy ($Q_\mathrm{R}/Q^\star_\mathrm{RD}\lesssim0.1$) as well as collisions with high impact energy ($Q_\mathrm{R}/Q^\star_\mathrm{RD}\gtrsim1.8$), as a shown in \cite{Housen1999}.
To address this issue, \citetalias{Leinhardt2012} included the second branch, referred to as super--catastrophic regime, that better models the linearity in the log-log space observed by various authors and summarised in \cite{Holsapple2002}.

The model for the catastrophic disruption criterion ($Q^\star_\mathrm{RD}$) in the gravity regime was derived by \cite{Housen1990} using $\pi$--scaling theory, and it was rearranged by \citetalias{Leinhardt2012} as follow:
\begin{equation}\label{eq:Qstar}
    Q_\mathrm{RD}^\star = c^* \frac{4}{5}\pi\rho_1GR^2_\mathrm{C1}
    \left[\frac{1}{4}\frac{(1+\gamma)^2}{\gamma}\right]^{-1+\left[2/(3\bar{\mu})\right]}\,,
\end{equation}
where $c^*$ is a scaling constant equivalent to the offset with respect the gravitational binding energy, $G$ is the gravitational constant, $R_\mathrm{C1}$ is the radius corresponding to a spherical object with mass $M_\mathrm{tot}$ and density \mbox{$\rho_1=1$ g/cm$^3$}, \mbox{$\gamma=M_\mathrm{p}/M_\mathrm{t}$} is the ratio between projectile and target mass, and $\bar{\mu}$ is a dimensionless material constant related to the energy and momentum coupling between projectile and target.

This model only works for head--on collisions.
When the collisional angle ($\theta$) exceeds the threshold value of \mbox{$\sin\theta_\mathrm{crit}=(R_\mathrm{t}-R_\mathrm{p})/(R_\mathrm{t}+R_\mathrm{p})$}, with $R_\mathrm{t}$ and $R_\mathrm{p}$ being the radius of the target and projectile respectively, not all the mass of the projectile interacts with the target during the collision.
Nevertheless, \citetalias{Leinhardt2012} model can still be applied to oblique impacts by considering the equivalent collision in which only the interacting mass of the projectile is employed.
{Due} to the extent of the new dataset, we decided to consider head--on only, i.e. all the collisions that satisfy $\theta<\theta_\mathrm{crit}$.

\subsubsection{LS12 scaling law - Original fits}\label{subsec:original}

\begin{figure*}[t]
  \centering
  \hspace*{-0.2cm} 
    \includegraphics[width=1.1\textwidth]{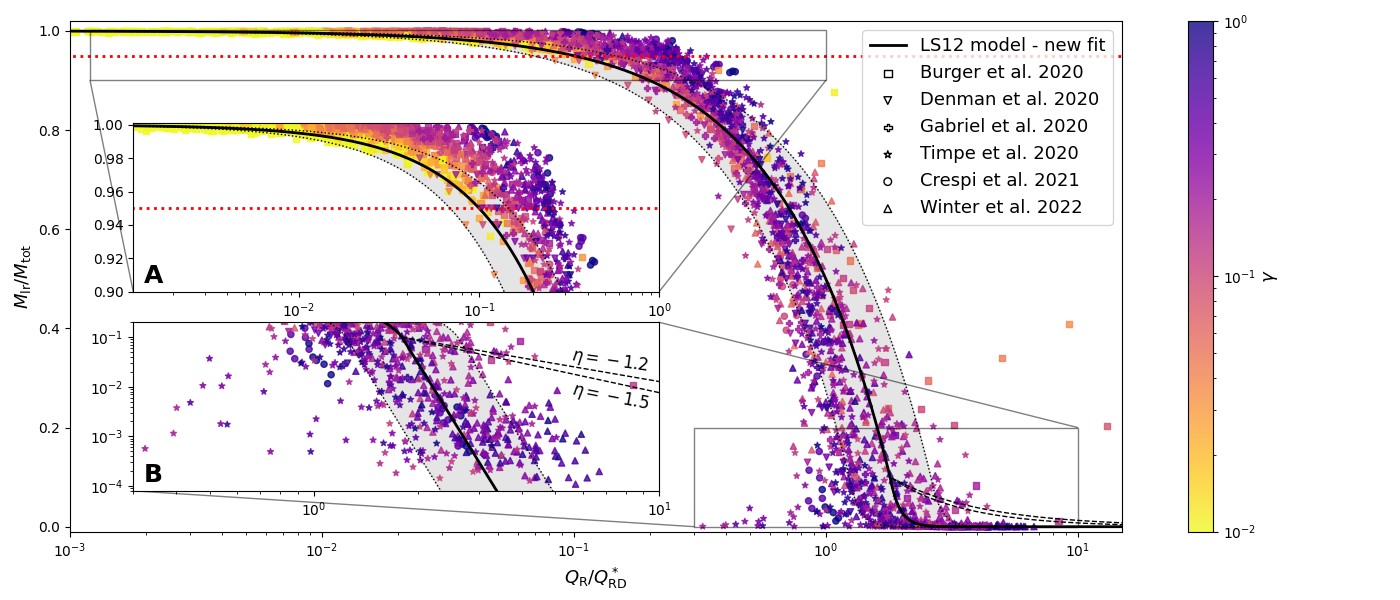}
  \caption{Scaled mass of the largest remnant ($M_\mathrm{lr}/M_\mathrm{tot}$) with respect to the impact energy scaled by the catastrophic disruption criterion ($Q_\mathrm{R}/Q^\star_\mathrm{RD}$). {The six datasets used in this study are represented with different symbols: square for data from \cite{Burger2020}, upside--down triangle for data from \cite{Denman2020}, plus symbol for data from \cite{Gabriel2020}, star symbold for \cite{Timpe2020}, circles for data from \cite{Crespi2021}, and triangles for data from \cite{Winter2022}.} The colors denote the mass ratio $\gamma$. The black line corresponds to the universal law {(Equation \ref{eq:universallaw})} and the gray area represent the dispersion $\delta$. The range of possible slope $\eta$ from \citetalias{Leinhardt2012} is represented with dashed lines in the super--catastrophic regime. The most crowded regions are zoomed in the panels A (merging regime) and B (super--catastrophic regime). The region above the red dotted line at $M_\mathrm{lr}/M_\mathrm{tot}=0.95$ is not included in the MCMC analysis.
  }\label{fig:LS12fit}
\end{figure*}

Equation \ref{eq:universallaw}, in conjunction with Equation \ref{eq:Qstar}, enables the estimation of the mass of the largest post--collisional remnant, given the impact energy ($Q^\star_\mathrm{RD}$) and the combined radius ($R_\mathrm{C1}$).
The model incorporates three free parameters: the scaling constant $c^*$, the material constant $\bar{\mu}$, and the slope $\eta$ for the super--catastrophic regime.

{In the original approach employed by \citetalias{Leinhardt2012}, they conducted three fits. First, using data from the works of \cite{Benz2007}, \cite{Marcus2009}, and \cite{Marcus2010}, they estimated the catastrophic disruption criterion $Q_\mathrm{RD}^\star$ for edge-on collisions through linear interpolation of simulations with similar collision parameters.
Subsequently, they fit Equation \ref{eq:Qstar} to the resulting distribution of $Q_\mathrm{RD}^\star$ as a function of $R_\mathrm{C1}$, obtaining the values $c^*=1.9\pm0.3$ and $\bar{\mu}=0.36\pm0.01$ for the model parameters.
Lastly, they estimated the slope $\eta$ from the data points in the supercathastrophic regime (second branch of Equation \ref{eq:universallaw}).
However, due to the limited number of data points in this regime, the parameter $\eta$ was not well--constrained. As a result, \citetalias{Leinhardt2012} recommended using the value $\eta = -1.5$ based on laboratory studies (\citealt{Kato1995} and \citealt{Fujiwara1977}).}


\subsubsection{{LS12 scaling law - New fit}}\label{subsec:original_but_new}

In contrast to the procedure implemented by \citetalias{Leinhardt2012} of {first estimating $c^\star$ and $\bar{\mu}$ from the  linear interpolation of $Q^\star_\mathrm{RD}$, and secondly evaluating the remaining model parameter $\eta$} given $Q_\mathrm{R}/Q_\mathrm{RD}^*$ and $M_\mathrm{lr}/M_\mathrm{tot}$, we decided to use the simulation data in its entirety ($Q_\mathrm{R}$, $R_\mathrm{C1}$, $\gamma$, $M_\mathrm{lr}/M_\mathrm{tot}$) to directly estimate all the model parameters ($c^*$, $\bar{\mu}$, $\eta$) in one go.
Furthermore, we introduced a new parameter $\delta$ that allows to estimate the data dispersion.
In particular, the dispersion is assumed to be related to the value $Q/Q_\mathrm{RD}^\star$ and is modelled, in log--space, by a normal distribution with constant standard deviation ($\delta$).
In other words, the measured value of $Q_\mathrm{R}/Q_\mathrm{RD}^\star$ is given by
\begin{equation}
    \log \left[\frac{Q_\mathrm{R}}{Q_\mathrm{RD}^\star}\right]_\mathrm{data} = 
    \log \left[\frac{Q_\mathrm{R}}{Q_\mathrm{RD}^\star}\right]_\mathrm{model} + \mathcal{N}(\mu=0;\,\delta)\,
\end{equation}
where $\mathcal{N}(\mu;\,\delta)$ is the normal distribution centered in $\mu$ with standard deviation $\delta$.

We performed a MCMC analysis with the aim of obtaining the posterior probabilities for the three model parameters plus $\delta$ as an additional free parameter.
Given the dispersion model, we assumed the likelihood ($\mathcal{L}$) to be defined as:
\begin{equation}\label{eq:like}
    \log \mathcal{L} = -\frac{N}{2}\log(2\pi\delta^2)
    + \sum_i
    \left[\frac{\log \left(\frac{\left[Q_\mathrm{R}/Q_\mathrm{RD}^\star\right]_i}{ 
    \left[Q_\mathrm{R}/Q_\mathrm{RD}^\star\right]_\mathrm{model}}\right)}{\delta}\right]^2\,,
\end{equation}
where the sum is over all the $N$ data, and $\left[Q_\mathrm{R}/Q_\mathrm{RD}^\star\right]_\mathrm{model}$ is obtained by inverting Equation \ref{eq:universallaw}.
We assumed uniform priors for all the free parameters in an wide interval around the values estimated by \citetalias{Leinhardt2012}.

We ran the MCMC analysis on the entire dataset and found that the solution is strongly biased by the collisions in the merging regime ($M_\mathrm{lr}/M_\mathrm{tot}\gtrsim0.9$).
Among all the head--on collisions, {more than a third} of them have \mbox{$M_\mathrm{lr}/M_\mathrm{tot}>0.95$}.
This unbalance in the dataset distribution cause the MCMC to converge on a solution that favours accurate modelling of the merging regime at the expense of the remaining dataset.
As evident from laboratory experiments (e.g. \citealt{Takagi1984}, \citealt{Housen1990}, \citealt{Nagaoka2014}, \citealt{Arakawa2022}), the model in Equation \ref{eq:universallaw} tend to underestimate the mass of the largest remnant at very low energy.
We therefore decided to exclude all the collisions with $M_\mathrm{lr}/M_\mathrm{tot}>0.95$ from the MCMC fitting procedure.

\begin{figure*}[t]
  \centering
  \hspace*{-0.2cm} 
    \includegraphics[width=1.08\textwidth]{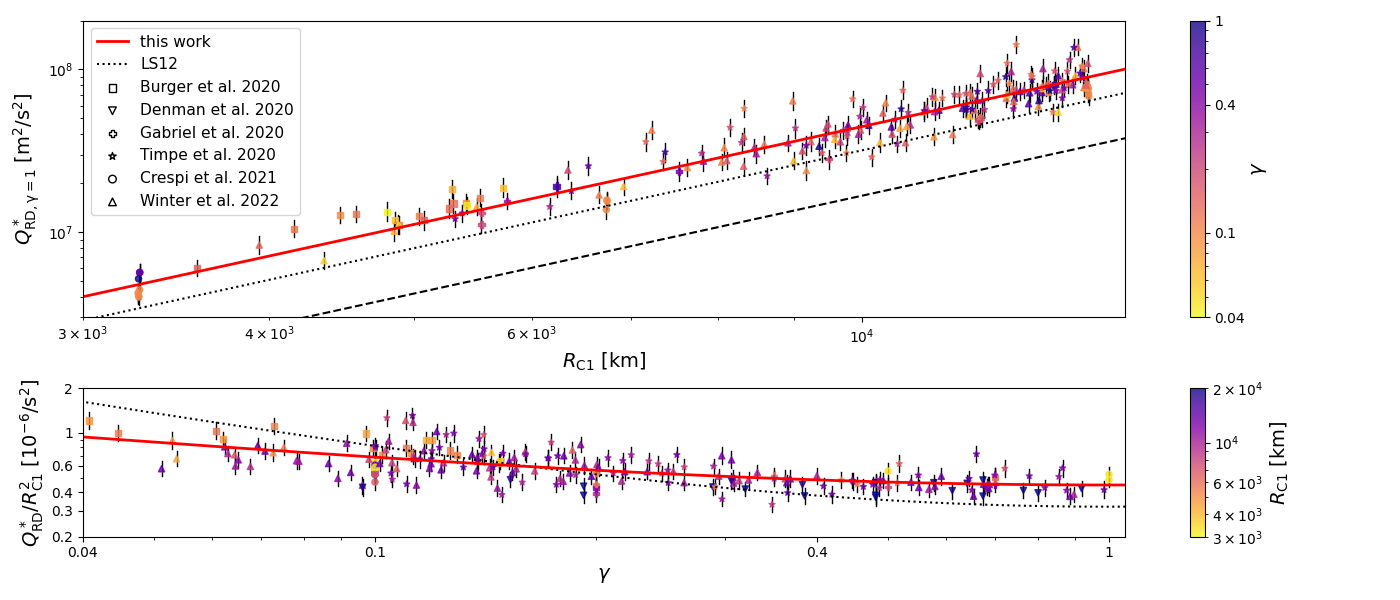}
  \caption{\textit{Top}: catastrophic disruption criterion for same--mass collisions ($Q^\star_\mathrm{RD,\gamma=1}$) with respect to the combined radius ($R_\mathrm{C1}$). The value $Q^\star_\mathrm{RD,\gamma=1}$ is obtained from Equation \ref{eq:Qstar} by dividing $Q^\star_\mathrm{RD}$ by the $\gamma$--dependent component $\left[(1+\gamma)^2/4\gamma\right]^{2/(3\bar{\mu})-1}$. \textit{Bottom}: catastrophic disruption criterion for same--mass collisions scaled by the combined radius squared ($Q^\star_\mathrm{RD,\gamma=1}/R_\mathrm{C1}^2$) with respect to the mass {ratio} ($\gamma$).  The symbols are the same as in Figure \ref{fig:LS12fit}. The colors denote the mass ratio $\gamma$ in the plot at the top, and the combined radius $R_\mathrm{C1}$ in the plot at the bottom. The solid red line corresponds to the best--fit from this work, i.e. {$c^*=2.661$ and $\bar{\mu}=0.4797$}, while the dotted black line corresponds to the best--fit from \citetalias{Leinhardt2012}, i.e. $c^*=1.9$ and $\bar{\mu}=0.36$. The binding energy (dashed black line) is shown for reference in the top panel.
  }\label{fig:Qstarfit}
\end{figure*}

{
We used uniform priors for the 4 model parameters, specifically $\mathcal{U}(0,100)$ for $c^*$, $\mathcal{U}(1/3,2/3)$ for $\bar{\mu}$,
$\mathcal{U}(-100,0)$ for $\eta$, and
$\mathcal{U}(0,2)$ for $\delta$. Here, $\mathcal{U}(a,b)$ represents the uniform distribution, with a density of $1/(b-a)$ in the interval $a \leq x < b$ and zero elsewhere.
The analysis of the posteriors gives the following results: $c^*=3.20\pm0.05$, \mbox{$\bar{\mu}=0.486\pm0.007$}, \mbox{$\eta=-11.4^{+0.7}_{-0.8}$}, and \mbox{$\delta=0.162\pm-0.003$}.
These values differ significantly from what estimated by \citetalias{Leinhardt2012}. In particular, the slope $\eta$ of the super--catastrophic regime deviates from the value $-1.5$ by more than 14 sigma.}

The value $\eta=-1.5$ suggested by \citetalias{Leinhardt2012} was derived from two laboratory studies of collisions between solid ice \citep{Kato1995} and collisions of polycarbonate projectiles against granite blocks \citep{Fujiwara1977}.
The strong discrepancy between these laboratory fragmentation experiments and the simulations in the datasets presented here could lie in the significantly different nature of the colliding bodies more than in the methodology (simulations versus laboratory experiments).
Collisions of ice and granite are in the strength regime, with the largest remnant being a single fragment of the largest body, while protoplanet collisions are in the gravity regime.
This different behavior is expected to be even more evident in the catastrophic regime, where most of the colliding mass is dispersed.

The datapoints derived through Equation \ref{eq:Qstar} and the best fit model of Equation \ref{eq:universallaw} are shown in \mbox{figure \ref{fig:LS12fit}}.
As expected, the universal law from \citetalias{Leinhardt2012} performs well for energies close to $Q^*_{RD}$, and, thanks to the new estimate of $\eta$, it also succeeds in predicting the collision outcome in the super--catastrophic regime (Figure \ref{fig:LS12fit} B).

A significant discrepancy between the \citetalias{Leinhardt2012} model and the simulations is still clearly present in the merging regime.
\citetalias{Leinhardt2012} scaling laws tend to overestimate the fragmentation efficiency for collisions with small impact energy ($Q_\mathrm{R}/Q_\mathrm{RD}^*\lesssim 0.3$), as shown in panel A of Figure \ref{fig:LS12fit}.
Moreover, we also noticed a strong correlation between this discrepancy and the mass ratio $\gamma$.

{In figure \ref{fig:LS12fit}, it is noticeable that there may be a dependence on the dataset, especially in the super-catastrophic regime (Panel B). Collisions simulated by \cite{Burger2020} and \cite{Winter2022} tend to cluster on the right-hand side of the best-fit result, indicating that more energy is required to break apart the colliding protoplanets. Conversely, collisions from the work of \cite{Timpe2020} exhibit the opposite trend.
To further investigate this potential behavior, we performed separate MCMC analyses for each dataset.}

{We found that the parameters $c^*$ and $\bar{\mu}$ are generally in agreement across the six datasets, typically differing by less than 2$\sigma$, with only a few exceptions. Notably, the value of $c^*$ obtained from the \cite{Crespi2021} dataset, $c^*=3.82\pm0.25$, exceeds the values obtained from the other datasets, which fall within the range of $c^*=3.03-3.40$. Additionally, the value of $\bar{\mu}$ obtained from the \cite{Denman2020} dataset, $\bar{\mu}=0.59\pm0.04$, surpasses the values of $\bar{\mu}=0.47^{+0.02}_{-0.01}$ and $\bar{\mu}=0.46\pm0.02$ obtained from the \cite{Timpe2020} and \cite{Burger2020} datasets, respectively.}

{We observed a bimodal behavior in the parameter $\eta$, with datasets yielding either extremely low values within the range of $-57$ to $-74$ and large errors, or datasets exhibiting high values of $\eta$ between -7 and -2 with smaller errors. Two prominent examples illustrating these behaviors are the \cite{Timpe2020} dataset, which yielded $\eta=-74^{+21}_{-18}$, and the \cite{Winter2022} dataset, resulting in $\eta=-7.0\pm0.3$. Both datasets are well-sampled within the super-catastrophic regime, each comprising more than 200 datapoints. However, the \cite{Timpe2020} dataset is concentrated around $M\mathrm{lr}/M_\mathrm{tot}\lesssim 0.1$, while the \cite{Winter2022} dataset is centered around $M_\mathrm{lr}/M_\mathrm{tot}\sim10^{-3}$.}

{Notably, the \citetalias{Leinhardt2012} model (equation \ref{eq:universallaw}) enforces the fit to pass through $M_\mathrm{lr}/M_\mathrm{tot}=0.1$ when $Q_\mathrm{R}/Q^*_\mathrm{RD}=1.8$, whereas the actual value is closer to $Q_\mathrm{R}/Q^*_\mathrm{RD}\sim1.3$. This discrepancy results in the significantly different estimates of $\eta$ for the \cite{Timpe2020} and \cite{Winter2022} datasets.}

{In addition, we observed varying degrees of dispersion among the different datasets, notably in simulations that incorporate the rotation of colliding protoplanets, as seen in the \cite{Timpe2020} and \cite{Winter2022} datasets. This dispersion is particularly noticeable in the super-catastrophic regime, where the presence of additional angular momentum can either aid or impede the dispersion of fragmented material.}

{Another distinguishing factor among the datasets arises from differences in the simulation routines and the composition of the colliding protoplanets. Nevertheless, these parameters appear to have a secondary influence compared to other collisional factors such as impact energy, masses involved, and impact angle. An exhaustive examination of how composition affects collision outcomes falls beyond the scope of this study.}

\subsubsection{New scaling law}\label{sec:new_model}

The need to model the $\gamma$--dependent offset between data and \citetalias{Leinhardt2012} scaling laws at low impact energies, together with the pursuit of a function that smoothly transitions from the merging regime to the log--log linear super--catastrophic regime {without fixing the transition point}, are two pivots around which we based the new model for the universal law.
A good model, able to satisfy these two requirements, is the product of an exponential function (for modelling the super--catastrophic regime) and a rational function (for modelling the merging regime).
This new version of the universal law is described by
\begin{equation}\label{eq:newModel}
    Q_\mathrm{R} = c_1Q_\mathrm{RD}^\star
    \left(2\frac{M_\mathrm{lr}}{M_\mathrm{tot}}\right)^{1/\eta}
    \frac{\left[1-\left(\frac{M_\mathrm{lr}}{M_\mathrm{tot}}\right)^{3/2}\right]^{\alpha(\gamma)}}
    {1+c_2\left(2\frac{M_\mathrm{lr}}{M_\mathrm{tot}}\right)^2}
    \, ,
\end{equation}
where $Q_\mathrm{RD}^\star$ can be obtained from Equation \ref{eq:Qstar}, $c_1$ and $c_2$ are constants and the exponent $\alpha$ is a function of the mass ratio $\gamma$.
We observed a linear dependence between the exponent $\alpha$ and $\log\gamma$.
Therefore, we decided to model $\alpha$ as $\alpha(\gamma)=\alpha_0+\sigma\cdot\log\gamma$, where $\alpha_0$ and $\sigma$ are two extra free parameters.
We also investigated the possibility for $\alpha$ to depend on the combined radius $R_\mathrm{C1}$ but no significant correlation was found.
We note that the new universal law is not analytically invertible and, in order to obtain $M_\mathrm{lr}/M_\mathrm{tot}$ given ($Q_\mathrm{R}$, $\gamma$, $R_\mathrm{C1}$), it would be necessarily to employ a simple root--finding algorithm.

The model in Equation \ref{eq:newModel} depends on three variables ($Q_\mathrm{R}$, $\gamma$, $R_\mathrm{C1}$) and six parameters, two of which ($c^*$, $\bar{\mu}$) arise from the physical model for $Q_\mathrm{RD}^\star$ (Equation \ref{eq:Qstar}) and the other five ($c_1$, $c_2$, $\eta$, $\alpha_0$, $\sigma$)  arise from the new analytical model (Equation \ref{eq:newModel}).
We decided to investigate these two sets of parameters separately so that the approximation inherent to the analytical model does not affect the estimate of the physical parameters $c^*$ and $\bar{\mu}$.

\begin{figure*}[t]
  \centering
  \hspace*{-0.2cm} 
    \includegraphics[width=1.1\textwidth]{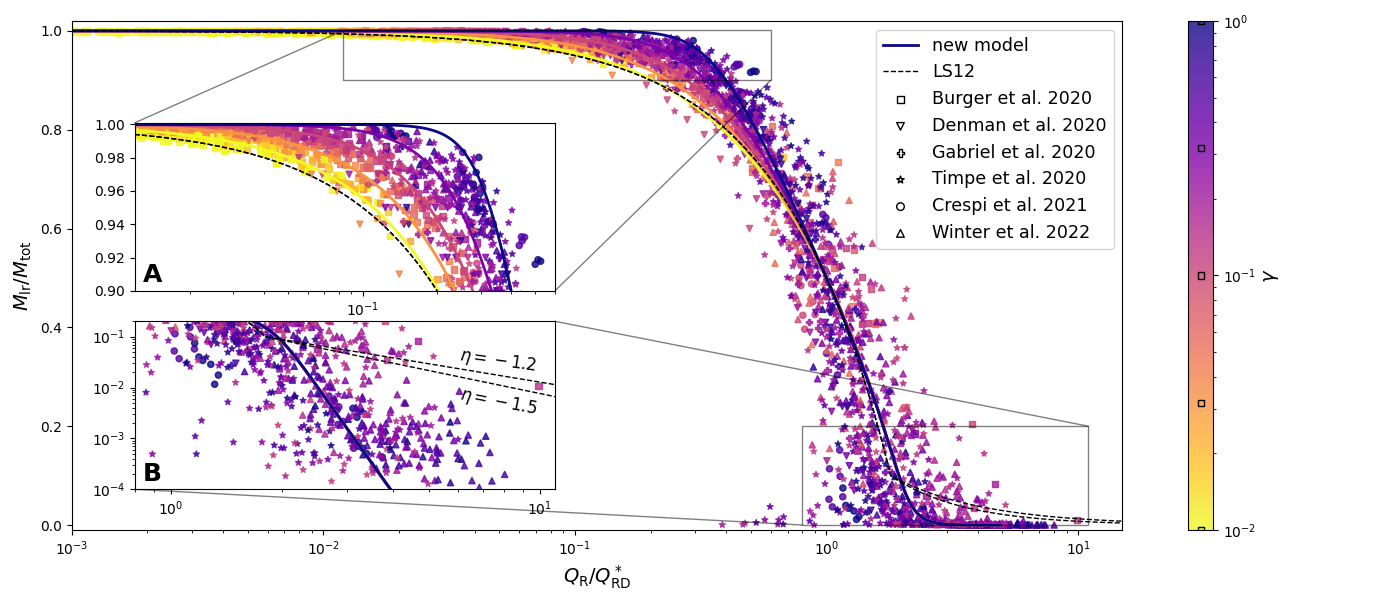}
  \caption{Scaled mass of the largest remnant ($M_\mathrm{lr}/M_\mathrm{tot}$) with respect to the impact energy scaled by the catastrophic disruption criterion ($Q_\mathrm{R}/Q^\star_\mathrm{RD}$). The symbols and the colors are the same as in Figure \ref{fig:LS12fit}. The coloured lines correspond to the new scaling law in Equation \ref{eq:newModel}, with each colour representing a different value for the mass ratio ($\gamma$), as indicated by the squares on the colorbar. The black dashed lines represent the original \citetalias{Leinhardt2012} universal law {(Section \ref{subsec:original})}. The most crowded regions are zoomed in the panels A (merging regime) and B (super--catastrophic regime).\vspace{4mm}
  }\label{fig:mymodel}
\end{figure*}

\begin{table*}[t]
  \centering
  \begin{tabular}{l|c|c|c|c|c|c}
  Model & $c^*$ & $\bar{\mu}$ & $\eta$ & $c_1$ & $\alpha_0$ & $\sigma$ \\
  \hline
  \hline
    \citetalias{Leinhardt2012} & $1.9\pm0.3$ & $0.36\pm0.01$ & -1.5 & - & - & - \\
  \hline
    \citetalias{Leinhardt2012} - new fit & $3.20\pm0.05$ & $0.486\pm0.007$ & $-11.4^{+0.7}_{-0.8}$ & - & - & - \\
  \hline
    new model & $2.66\pm0.04$ & $0.480\pm0.006$ & $-10.18\pm0.02$ & $1.707\pm0.001$ & $0.1755\pm0.0004$ & $-0.3252\pm0.0002$ \\
  \end{tabular}
  \vspace{3mm}
  \caption{Best fit parameters and associated error. \textit{First line}: fit of the \citetalias{Leinhardt2012} scaling law to the original dataset presented in \citetalias{Leinhardt2012} (values from \citetalias{Leinhardt2012}). \textit{Second line}: fit of the \citetalias{Leinhardt2012} scaling law to the dataset presented in this work. \textit{Third line}: fit of the new scaling law (Equation \ref{eq:newModel}) to the dataset presented in this work.}
  \label{tab:2}
  \vspace{-6mm}
\end{table*}

To obtain the catastrophic disruption criterion ($Q^\star_\mathrm{RD}$), we selected all the collisions with $M_\mathrm{lr}/M_\mathrm{tot}$ in the range {\mbox{$0.4-0.6$}}.
In this neighbourhood, $M_\mathrm{lr}/M_\mathrm{tot}$ scales linearly with the logarithm of the impact energy $\log Q_\mathrm{R}$ with slope -0.97, which has been obtained by fitting the datapoints with a linear function in the semilogarithmic space.
We used this linear relation to predict $Q^\star_\mathrm{RD}$ for each collision by assuming
\mbox{$\log Q^\star_\mathrm{RD} = \log Q_\mathrm{R} - 0.97\cdot\left(M_\mathrm{lr}/M_\mathrm{tot}-0.5\right)$}.
We note that, on average, the estimate of $Q^\star_\mathrm{RD}$ is not affected by the chosen value for the slope since the data are homogeneously distributed around $M_\mathrm{lr}/M_\mathrm{tot}=0.5$.
In other words, a different choice of slope would only increase (or decrease) the dispersion of $Q^\star_\mathrm{RD}$ around the true value.


Finally, we used the derived values $Q^\star_\mathrm{RD}$ in the functions of $R_\mathrm{C1}$ and $\gamma$ to determine the parameters $c^*$ and $\bar{\mu}$ from \mbox{Equation \ref{eq:Qstar}}.
The analysis of the posteriors yields the following results: {\mbox{$c^*=2.661^{+0.037}_{-0.036}$}, and \mbox{$\bar{\mu}=0.4797^{+0.0061}_{-0.0059}$}, with priors set to $\mathcal{U}(0,100)$ for $c^*$ and $\mathcal{U}(1/3,2/3)$ for $\bar{\mu}$}.
The data and the best-fit model are presented in Figure \ref{fig:Qstarfit}, alongside the results from \citetalias{Leinhardt2012} for comparison.


The new universal law (Equation \ref{eq:newModel}) depends on 5 model parameters. However, the degrees of freedom can be reduced to 4 by imposing \mbox{$Q^*_\mathrm{RD}=Q_\mathrm{R}|_{M_\mathrm{lr}/M_\mathrm{tot}=0.5}$}.
Consequently, we can rewrite $c_2$ as:
\begin{equation}
    c_2 = c_1\left(1-2^{-3/2}\right)^{\alpha(\gamma)}-1\, .
\end{equation}
To obtain the remaining 4 parameters we run an MCMC algorithm where we assumed {$c^*=2.661$ and \mbox{$\bar{\mu}=0.4797$}}.
We adopted the likelihood in Equation \ref{eq:like} where $\delta$ is derived by propagating the errors on $c^*$ and $\bar{\mu}$.
{The best-fit result is displayed in Figure \ref{fig:mymodel}, while the analysis of the posteriors yields the following results: \mbox{$c_1=1.7074^{+0.0011}_{-0.0012}$}, \mbox{$\eta=-10.179^{+0.021}_{-0.022}$}, \mbox{$\alpha_0=0.1754729^{+0.00038}_{-0.00039}$}, \mbox{$\sigma=-0.32516^{+0.00020}_{-0.00019}$}}. 
Moreover, we estimated the data dispersion along $\log Q_\mathrm{R}/Q^*_\mathrm{RD}$ and we obtained an approximately constant and symmetric dispersion of {0.11}.
The dispersion is attributed to various factors not considered in our model, including, but not limited to, the chemical composition of the colliding bodies, their rotation, and the angular momentum of the collision.
{For the MCMC analysis we used the following priors for the fit parameters $\mathcal{U}(0,100)$ for $c_1$, $\mathcal{U}(-100,0)$ for $\eta$, $\mathcal{U}(0,10)$ for $\alpha_0$, and $\mathcal{U}(-10,0)$ for $\sigma$.}

We noted that the new universal law is asymptotic for $M_\mathrm{lr}/M_\mathrm{tot}\rightarrow 1$. This behaviour is nonphysical and approximations must be employed.
From the MCMC analysis, we obtained that our model starts deviating from the measured values of $Q_\mathrm{R}/Q^*_\mathrm{RD}$ when \mbox{$M_\mathrm{lr}/M_\mathrm{tot}>0.999$}.
We suggest to assume the collisions in this regime to be the perfect merging of the two bodies \mbox{($M_\mathrm{lr}/M_\mathrm{tot}=1$)}.

{Following the approach outlined in the previous section, we conducted a separate analysis of the datasets. Generally, the results from different datasets exhibit consistency, and any observed discrepancies can be attributed to the differences between the datasets as described in Section \ref{subsec:original_but_new}. The most notable deviation was observed in the dataset from \cite{Denman2020}. In the case of this specific dataset, we observed that our model tends to overestimate the mass of the largest remnant during merging events (panel A of Figure \ref{fig:mymodel}). This deviation can be attributed to the presence of an atmosphere in the \cite{Denman2020} dataset, a feature absent in the other datasets we considered}

{
As noted by \cite{Denman2020}, low-energy impacts primarily result in atmosphere loss, while more energetic impacts are required to fragment both the mantle and the core of the colliding bodies. Therefore, caution should be exercised when applying our model to collisions involving planets with a substantial atmosphere, as it may not accurately represent the outcomes in such scenarios.}

\begin{figure*}[t]
  \centering
  \hspace*{-0.2cm} 
    \includegraphics[width=1.1\textwidth]{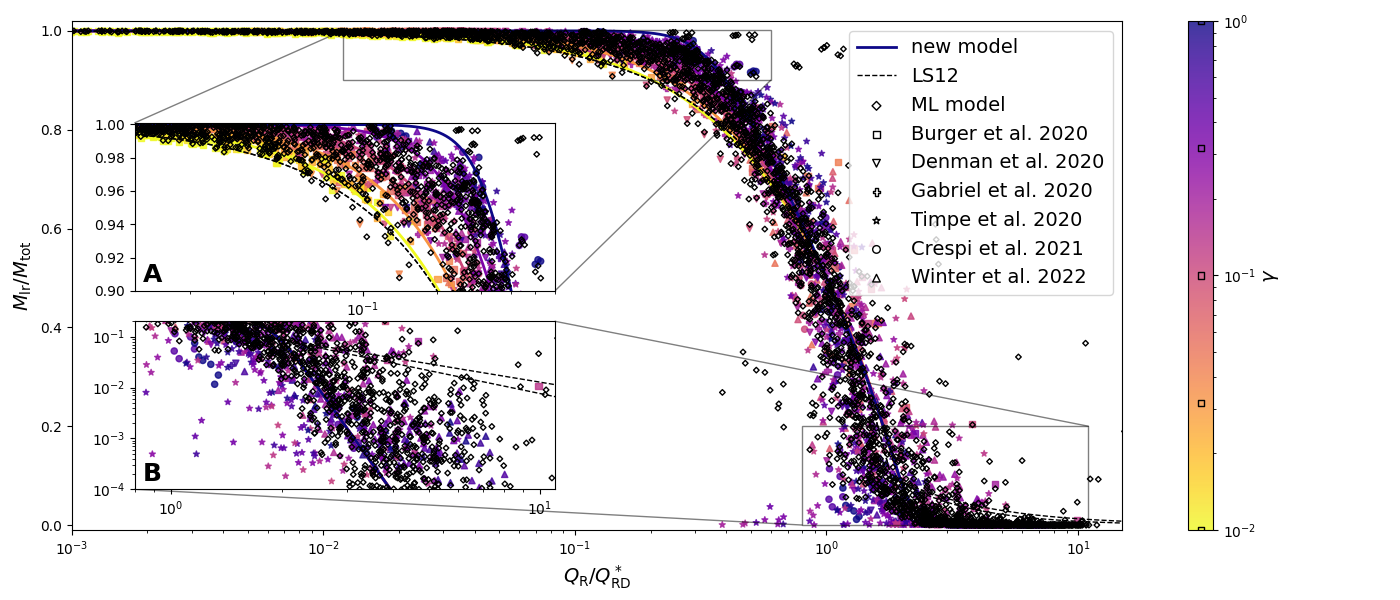}
  \caption{Scaled mass of the largest remnant ($M_\mathrm{lr}/M_\mathrm{tot}$) with respect to the impact energy scaled by the catastrophic disruption criterion ($Q_\mathrm{R}/Q^\star_\mathrm{RD}$). The empty diamonds represent the scaled mass predicted by the ML model. Coloured symbols and lines are the same as in Figure \ref{fig:mymodel}. The black dashed lines represent the original \citetalias{Leinhardt2012} universal law {(Section \ref{subsec:original})}.}\label{fig:comparison}
\end{figure*}

\subsection{Models comparison}
\label{restrictedprob}
Here we compare the performance of 3 different models: the analytical model from \citetalias{Leinhardt2012} with the new best fit parameters, the new analytical model presented in this work, and a simple Machine Learning (ML) model to provide a benchmark against which to compare our new analytical model. We thus trained a classic \texttt{Random Forest Regressor} \citep{scikit-learn} using its default hyperparameters and 3 features only: the impact energy ($Q_\mathrm{R}$), the combined radius ($R_\mathrm{C1}$), and the mass ratio ($\gamma$). We also limited the dataset to the head-on collision cases. 

{
The predicted values from the ML model, as well as the original data, are depicted in Figure \ref{fig:comparison}, compared to the original model from LS12 (Section \ref{fig:comparison}) and the new model introduced in this study (Section \ref{sec:new_model}).}

{
We observe that the ML model effectively captures the dispersion around the mean of the predicted quantity, a characteristic not readily attainable with analytical models. However, it's worth noting that in the super catastrophic regime, we observe deviations in the predicted outcomes from the ML model, especially in cases involving high rotators from the Timpe et al. (2020) dataset. This discrepancy may be attributed to the limited number of parameters on which the ML model has been trained.}

{
Introducing rotation as a parameter in the ML model could potentially enhance its predictive efficiency. Nevertheless, this falls outside the scope of our current study, which is primarily focused on evaluating and comparing the predictive capabilities of our analytical model against those of a ML model that operates without relying on analytical assumptions.}

{Quantitatively, metric scores for the different models are summarized in table \ref{metrics}. We calculate the root mean squared error  $\text{RMSE}(y, \hat{y}) = \sqrt{\frac{\sum_{i=0}^{N - 1} (y_i - \hat{y}_i)^2}{N}}$, median absolute error: $$\text{Med. Abs Err}(y, \hat{y}) = \text{median}(\mid y_1 - \hat{y}_1 \mid, \ldots, \mid y_n - \hat{y}_n \mid)$$ and median relative error:  $$\text{Med. Rel Err}(y, \hat{y}) = \text{median}(\mid y_1 - \hat{y}_1 \mid / \mid y_1 \mid , \ldots, \mid y_n - \hat{y}_n \mid / \mid y_n \mid)$$  between the actual and predicted scaled mass of the largest remnant.}
We find that while the 3 models have comparable root mean squared errors as this metric is dominated by large $M_\mathrm{lr}/M_\mathrm{tot}$ values, the new analytical model outperforms \citetalias{Leinhardt2012} by factors of respectively $\sim$ 6 and 4 on the more sensitive median absolute and relative errors. The errors of the ML model, taken as the average of a 10-fold cross validation, are all very close to our new analytical model, reflecting the fact that complex models are not needed for simple low dimensionality problems.

Retraining the model using all the available parameters, such as the mass the composition and the spin of the colliding bodies, did not significantly improve its overall performance for head-on cases.
This result confirms that the collision outcome is strongly dependent on the 3 parameters used in the scaling laws.

\begin{table}[h]
\begin{center}
\begin{tabular}{l|l|l|l}
\multicolumn{1}{l|}{Metric} &
  \multicolumn{1}{l|}{LS12} &
  \multicolumn{1}{l|}{new model} &
  \multicolumn{1}{l}{ML}\\ \hline
  \hline
RMSE           & 0.111 & 0.116  & 0.11 \\
Med. Abs. Err. & 0.0223 & 0.0054 & 0.004   \\
Med. Rel. Err. & 0.158  & 0.059 & 0.098\\
\end{tabular}
\vspace{3mm}
\caption{Accuracy metrics to compare the analytical models of \citetalias{Leinhardt2012} and this work, in addition to our restricted ML model. RMSE is the root mean squared error, Med. Abs Err is the median absolute error, and  Med. Rel Err is the median relative error. }
\label{metrics}
\end{center}
\vspace{-6mm}
\end{table}

\section{Summary \& conclusions} \label{sec:summary}

\subsection{LS12 scaling laws compared to new data}
In this work, we reviewed the two main analytical models upon which the widely used scaling laws from \cite{Leinhardt2012} are founded, precisely: the catastrophic disruption criterion, which allows us to estimate the collision energy needed to disperse half of the total mass involved in the collision, and the universal law, which allows us to predict the mass of the largest post--collisional remnant ($M_\mathrm{lr}/M_\mathrm{tot}$).
{We used six datasets of SPH simulations of collisions from the works of \cite{Burger2020}, \cite{Denman2020}, \cite{Gabriel2020}, \cite{Timpe2020}, \cite{Crespi2021}, and \cite{Winter2022}, for a total of more than 32000 simulations}.

By comparing the \citetalias{Leinhardt2012} scaling laws with the new datasets we observed that these laws tend to underestimate the mass of the largest remnant in the accretion regime ($M_\mathrm{lr}/M_\mathrm{tot}\gtrsim0.9$).
In this regime, we also noticed a strong dependence between $M_\mathrm{lr}/M_\mathrm{tot}$ and the mass ratio of the colliding bodies ($\gamma$). In particular, collisions with the same scaled impact energy ($Q_\mathrm{R}/Q^\star_\mathrm{RD}$) but a smaller mass ratio of the colliding bodies tend to result in a less efficient accretion/merger than collisions with a larger mass ratio.
In the catastrophic regime ($M_\mathrm{lr}/M_\mathrm{tot}\gtrsim0.9$), we observed a strong discrepancy between \citetalias{Leinhardt2012} scaling laws and our dataset.
In particular, we obtained a slope of {$\eta=-11.4^{+0.7}_{-0.8}$} when fitting the \citetalias{Leinhardt2012} model to our dataset, compared to $\eta=-1.2\sim-1.5$ predicted by \citetalias{Leinhardt2012}.

\subsection{New scaling laws}
We developed an analytical scaling law that, analogously to the universal law from \citetalias{Leinhardt2012}, can be used to predict the mass of the largest remnant of a collision between gravity--dominated bodies. Our model (Equation \ref{eq:newModel}) is able to reproduce the $\gamma$--dependent distribution observed in the accretive regime, as well as exponential decrease in the catastrophic regime. It is valid for \mbox{$M_\mathrm{lr}/M_\mathrm{tot}<0.999$}, beyond which we  suggest assuming that the collision resulted in an inelastic merger.
Following the work of \citetalias{Leinhardt2012}, we assumed the catastrophic disruption criterion in the gravity regime to be modeled by Equation \ref{eq:Qstar}, and we found best--fit parameters {\mbox{$c^*=2.661^{+0.037}_{-0.036}$}, and \mbox{$\bar{\mu}=0.4797^{+0.0061}_{-0.0059}$}}.
\citetalias{Leinhardt2012} estimated these two parameters to be $c^*=1.9\pm0.3$, \mbox{$\bar{\mu}=0.36\pm0.01$}.
Our estimate for the offset parameter $c^*$ is at the border of compatibility with what was obtained by \citetalias{Leinhardt2012}. However, it is interesting to notice that \citetalias{Leinhardt2012} estimate, by being 30\% smaller than what we observed, results in a more efficient fragmentation of the main colliding body and, therefore, an overestimation of the debris production.
The $\bar{\mu}$ value obtained by \citetalias{Leinhardt2012} is indicative of almost pure momentum scaling for gravity--dominated bodies, while our value suggests a balanced combination between momentum and energy coupling. However, caution must be practiced when deriving any significant physical conclusion about the energy--momentum coupling since both \citetalias{Leinhardt2012} and our estimate of $\bar{\mu}$ fit well inside the data dispersion (see Figure \ref{fig:Qstarfit}). Finally, we found that  ML models such as a \texttt{Random Forest Regressor} does not perform better than the new analytical model, confirming the prediction efficiency of the latter.


\section*{Acknowledgments}

We would like to express our gratitude to Christoph Sch\"{a}fer for his valuable comments and feedback on this manuscript. His suggestions and criticism have greatly improved the quality of our work. This material is based upon work supported by Tamkeen under the NYU Abu Dhabi Research Institute grant CASS.

\bibliographystyle{aa}
\bibliography{refs.bib}

\end{document}